\documentclass[aps,prb,reprint,showpacs,amsmath,amssymb,floatfix,superscriptaddress]{revtex4-1}
\usepackage{color}
\usepackage{fancyhdr}
\usepackage{amssymb,latexsym}
\usepackage{amsmath}
\usepackage[latin1]{inputenc}
\usepackage{epsfig}
\usepackage{eucal}
\usepackage{amsfonts}
\usepackage{graphicx}
\usepackage{subfigure}
\begin{document}

\title{The Interplay of Charge and Spin in Quantum Dots: The Ising Case}
\author{Boaz Nissan-Cohen}
\author{Yuval Gefen}
\affiliation{Department of Condensed Matter Physics, The Weizmann Institute of Science, Rehovot 76100, Israel}
\author{Mikhail Kiselev}
\affiliation{The Abdus Salam International Centre for Theoretical
Physics, Strada Costiera 11, I-34151 Trieste, Italy}
\author{Igor Lerner}
\affiliation{School of Physics and Astronomy, University of Birmingham, Birmingham B15 2TT, UK}

\date{\today}

\begin{abstract}
The physics of quantum dots  is succinctly depicted by the
{\it Universal Hamiltonian}, where only zero mode interactions are
included. In the case where the latter involve charging
and isotropic spin-exchange terms,  this would lead to
a non-Abelian action. Here we address an Ising
spin-exchange interaction, which leads to an Abelian action. The
analysis of this simplified yet non-trivial model shed some light on a more general case of charge and spin entanglement. We present a
calculation of the tunneling density
of states and of the dynamic magnetic susceptibility. Our results  are
amenable to experimental study and may allow for an experimental
determination of the exchange interaction  strength.
\end{abstract}
\pacs{73.23.Hk,73.63.Kv,75.75.+a,75.30.Gw}



\maketitle

\section{Introduction}
\label{intro} Significant progress in the study of the physics of
quantum dots (QDs) has been achieved following the introduction of
the {\it Universal Hamiltonian}\cite{kurland1992mmf,ABG} (UH).
The latter facilitated the simplification of intricate
electron-electron interactions within a QD in a controlled way.
Within that scheme interactions are represented as the sum of three
spatially independent terms: charging, spin-exchange, and Cooper
channel. Notably, even the inclusion of the first two terms turned
out to be non-trivial: the resulting action is
non-Abelian\cite{kiselev2006isa}.

To understand the complexity of such a problem one can refer to the
case of charging-only interaction. As was suggested by Kamenev and
Gefen\cite{kamenev1996zba}, one can take the following steps in
solving that problem: start from a fermionic action which includes
an interaction term quadric in the (fermionic Grassman) variables,
perform a Hubbard-Stratonovich transformation by introducing an
auxiliary bosonic field, then perform a gauge transformation over
the Grassman variables, and finally integrate them out. The
resulting, purely bosonic, action is simple. In an imaginary time
(Matsubara) picture the action is quadratic in the  bosonic components, which renders this action easily
solvable. The trick of gauge-integrating over Grassman variables does not
work for the non-Abelian case\cite{kiselev2006isa} so that an alternative approach
is  needed.

Attempts to account for charge and spin interactions in QD have been
reported earlier. Alhassid and Rupp \cite{alhassid2003esa} have
found an exact solution for the partition function (and
susceptibility); elements of their analysis were then incorporated
in a master equation analysis of transport through the QD. More
recently an exact solution of the isotropic spin interaction model
has been presented\cite{burmistrov,*burmistrov2}. For the latter model some
quantities turn out to be particularly simple (e.g. the finite
frequency spin susceptibility vanishes; evidently there is no
difference between longitudinal and transverse spin susceptibility).
This means that the analysis of a model with anisotropy in the spin
interaction is called for. A perturbation expansion in spin
anisotropy has been reported earlier\cite{kiselev2006isa}, but it
still remains desirable to consider an anisotropic model which can
be analyzed exactly. By considering such a model one would be able
to understand the entanglement between charge and spin degrees of
freedom, and also see in detail how a non-vanishing, complex spin
susceptibility arises. This is the focal point of the present
analysis.

In bulk systems the exchange interaction competes with the kinetic
energy leading to Stoner Instability (SI). \cite{Stoner} In
finite size systems mesoscopic Stoner unstable regime may be a
precursor of bulk thermodynamic SI. We consider here an Ising spin
interaction. Such a model is Abelian, and complications due to
non-commutativity of different terms in the action do not arise
here. Also such a model does not exhibit a mesoscopic Stoner
unstable regime \cite{ABG}. This means that at zero
temperature, as the dimensionless parameter $J/\Delta$ ($J$ being
the exchange interaction strength and $\Delta$ is the mean level
spacing) the system   abruptly switches from a paramagnetic to a
(thermodynamic Stoner unstable) ferromagnetic phase. We stress that
notwithstanding the simplicity of the model considered, spin-charge
entanglement is present here, and non-trivial transverse a.c.\
susceptibility does arise. Some of our conclusions can in principle
tested in QDs made of materials close to the
thermodynamic Stoner Instability, e.g., Co impurities in Pd or Pt
host, Fe or Mn dissolved in various transition metal alloys, Ni impurities
in Pd host, and Co in Fe grains, as well as new nearly ferromagnetic
rare earth materials.%
\cite{Exp:Co_in_Pt,*Exp:Co_in_Pt-1,*Exp:Co_in_Pt-2,%
Exp:Fe_in_TransMet,*Exp:Ni_in_Pd,*Mirza,Menon,*Canfield}

The outline of this paper is as follows. In Section II we introduce
our model Hamiltonian and the subsequent imaginary time action. In
Section III we employ the technique of zero-dimensional functional
bosonization,\cite{kiselev2006isa} which eventually allows us to express the
single-particle Green's Function as a product of the non-interacting
Green's Function and a term which depends on two bosonic fields. We then show how to reduce the problem to that of
classical stochastic equations for the bosonic fields.   In section IV we express the
grand-canonical partition function in terms of canonical ones,
leading to both a mathematical and physical simplification of the
calculation. In Section V we calculate the tunneling density of states  and  in Section VI longitudinal and transverse spin
susceptibilities. Section VII presents a
{summary of the main results} with some perspectives. We include some more technical calculations in three Appendices.

\section{Hamiltonian and Effective Action}
\label{model} We consider a normal-metal  QD in the metallic regime, where the Thouless energy $E_{\mathrm{Th}}$ and the mean level
spacing $\Delta$ satisfy $g\equiv E_{\mathrm{Th}}/\Delta\gg1$ (g is the dimensionless
conductance) and a temperature $T\gg\Delta$. It is the regime where a
description in terms of UH is
viable.

We restrict ourselves to a simplified version of the UH where   the interaction in the Cooper channel is set to zero and the
 spin-exchange term is chosen to be a fully anisotropic
Ising-like term,  $-J\hat{S}_Z^2$, with a
ferromagnetic exchange coupling, $J>0$, $\hat{S}_Z$ is the total spin of the dot in the
$\hat{z}$ direction. This form of
interaction is sufficient to bring about the Stoner instability
phenomenon and   other   spin-related effects,
whilst avoiding  calculational complexities inherent to a fully
spin-symmetric model. Possible physical sources for such an
anisotropy may include geometrical and/or molecular anisotropy,
magnetic impurities in the system, or even the application of
anisotropic mechanical pressure.

The complete form of the reduced UH is thus
\begin{align}\label{H}
\nonumber
H=&\sum_{\alpha,\sigma}\varepsilon_\alpha a^\dag_{\alpha,\sigma}
a_{\alpha,\sigma}+E_{\text{c}} \Big[\sum_{\alpha,\sigma}a^\dag_{\alpha,\sigma}
a_{\alpha,\sigma}-N_0\Big]^2\\
&-\frac{J}{4}\Big[\sum_{\alpha}a^\dag_{\alpha,\sigma}
\sigma^z_{\sigma\sigma'}a_{\alpha,\sigma'}\Big]^2.
\end{align}
Here $\{{\varepsilon _\alpha}\}$ is a set of  electronic levels in the dot, and $N_0$ in the charging term  represents
a positive background charge  controlled via an
external gate. We assume that the QD is either isolated or weakly-coupled to the leads and in the Coulomb blockade regime. On the other hand, we will be  considering  the spin-disordered
regime  below the Stoner instability. So the parameters of the Hamiltonian (\ref{H})  obey
 \begin{align}\label{ineq}
    J<\Delta\ll T\ll E_{\text{c}}
\,, \end{align}
  where  $T\equiv\beta^{-1}$ is the  temperature.

The Euclidean action corresponding to the Hamiltonian  (\ref{H}) is given by
\begin{align}\label{eq:action1}
\nonumber S[&\overline{\Psi} ,\Psi ]=\sum_\alpha\int_0^\beta
\mathrm{d}\tau\bigg\{\overline{\Psi}_\alpha
(\partial_\tau+\varepsilon_\alpha-\mu)\Psi_\alpha \\ &+E_{\text{c}} \Big[\sum_\alpha\overline{\Psi}_\alpha\Psi_\alpha-N_0\Big]^2
-\frac{J}{4}\Big[\sum_\alpha\overline{\Psi}_\alpha\sigma^z\Psi_\alpha\Big]^2 \bigg\},
\end{align}
where we use spinor notations
$\overline{\Psi}_\alpha=\left(\overline{\psi}_{\uparrow\alpha}(\tau), \overline{\psi}_{\downarrow\alpha}(\tau)\right)$. We
introduce two auxiliary bosonic fields, $\varphi^{{\text c}}(\tau)$ and
$\varphi^{{\text s}}(\tau)$, to decouple the Coulomb and exchange terms with the help of
 a standard Hubbard-Stratonovich (HS) transformation. This results in the following action:
\begin{equation}\label{eq:action2}
S=S ^{{\text c}} +S  ^{{\text s}} +S^{{\text{mix}} },
\end{equation}
where
     \begin{align}\notag
        &S ^{{\text c}} =\int_0^\beta \!\!\mathrm{d}\tau\left[\frac{\varphi^{{\text c}}(\tau)^2}{4E_{\text{c}} }-
        \mathrm{i}N_0\varphi^{{\text c}}(\tau)\right] \,,
    \\\label{eq:actions}
        &S^{{\text s}}=\int_0^\beta \!\!\mathrm{d}\tau\frac{\varphi^{{\text s}}(\tau)^2}{J} \,,
    \\\notag
        &S^{{\text{mix}} }=\int_0^\beta \!\!\mathrm{d}\tau\sum_\alpha\overline{\Psi}_\alpha
        \left[\partial_\tau+\varepsilon _\alpha-\mu+\mathrm{i} \varphi^{{\text c}} +\sigma^z\varphi^{{\text s}}
        \right]\Psi_\alpha.
    \end{align}
Here $\sigma^z$ is a Pauli
matrix, the bosonic fields are periodic and the fermionic fields are antiperiodic in $\tau$ with period $\beta$.  This  action is the starting point for all the subsequent calculations.  We will use the functional bosonization approach as developed in \cite{Grishin,kiselev2006isa}: first we gauge out the mixed fermionic-bosonic terms in the action (\ref{eq:actions}) and then integrate over the fermionic field thus arriving at a purely bosonic action. After that, instead of dealing with this action directly we will use a stochastic bosonization as described in the following section.

\section{From Functional to Stochastic Bosonization}
\label{sec:ftos}

In order to gauge out the mixed fermionic-bosonic terms in the action (\ref{eq:actions}), we introduce a generalized gauge
transformation, $
\widetilde{\Psi}_\alpha= \mathcal{T}^{-1}\Psi_\alpha \,,\;
\widetilde{\overline{\Psi}}_\alpha=\overline{\Psi}_\alpha  \mathcal{T}  $ with
\begin{equation}\notag
 \mathcal{T}=\mathrm{e}^{\mathrm{i} \theta^{{\text c}} (\tau)\mathcal{I}+\theta^{{\text s}} (\tau)\sigma^z}= \left(\begin{array}{cc}
\mathrm{e}^{\mathrm{i} \theta^{{\text c}} (\tau)+\theta^{{\text s}} (\tau)} & 0 \\
0 & \mathrm{e}^{\mathrm{i} \theta^{{\text c}} (\tau)-\theta^{{\text s}} (\tau)} \\
\end{array}\right)\,.
\end{equation}
``Gauging out'' implies the following identity
\begin{align}\label{GO}
    \overline{\Psi}_\alpha\left[\partial_\tau+\mathrm{i} \varphi^{{\text c}}(\tau)+\sigma^z\varphi^{{\text s}}(\tau)\right]\Psi_\alpha=
\widetilde{\overline{\Psi}}_\alpha \left[\partial_\tau+\mathcal{A}\right]\widetilde{\Psi}_\alpha\,,
\end{align}
where $\mathcal{A}$ is some constant matrix. In order to fulfill (\ref{GO}) we require the gauge matrix $ \mathcal{T}$ to obey
\begin{equation}\label{eq:gaugeequation}
\left[\partial_\tau+\mathrm{i} \varphi^{{\mathrm c}}(\tau)+\sigma^z\varphi^{{\text s}}(\tau)\right] \mathcal{T}= \mathcal{TA}\,.
\end{equation}
  Since
the bosonic fields are real, this equation
 separates into real and imaginary parts,
corresponding to the exchange and charge channels. Using the substitution $\mathcal{A}=\mathcal{A}^{{\text s}} \sigma^z+{\mathrm{i}} \mathcal{A}^{{\text c}} $
 for the constant matrix ${\mathcal{A}}$ in
the matrix gauge equation (\ref{eq:gaugeequation}), we have
    \begin{align}\label{theta}
        \dot{\theta}^a (\tau)&=\mathcal{A}^a  -\varphi^a(\tau)
    \end{align}
where $a$ stands either for charge, c, or for spin, s.

To determine  the constants $\mathcal{A}^{{\text s}} $ and $\mathcal{A}^{{\text c}} $ we note
that  the antiperiodicity of the fermionic
fields  requires that $ \mathcal{T}(\beta)= \mathcal{T}(0)$. This in turn implies
$\theta^{{\text s}} (\beta)=\theta^{{\text s}} (0)+2\pi {\mathrm{i}}  n^{{\text s}} $ and
$\theta^{{\text c}} (\beta)=\theta^{{\text c}} (0)+2\pi N  $ with integer $n^{{\text s}} $ and $N  $.
Now we single out zero-Matsubara-frequency components of the bosonic fields $\varphi^a(\tau)$:
    \begin{align}\label{phi0}
        \varphi^a(\tau)&=\varphi^a_0+\widetilde{\varphi}^a(\tau) \,,& \beta\varphi^a _0&\equiv \int_0^\beta\!\!\mathrm{d}\tau\varphi^a(\tau)\,.
    \end{align}
Integrating Eqs.~(\ref{theta}) over $\tau$ from $0$ to $\beta$ results in   $ \mathcal{A}^{{\text c}} =\varphi^{{\text c}}_0+({{2\pi}/{\beta}})N $  and $ \mathcal{A}^{{\text s}} =\varphi^{{\text s}}_0+({{2\pi{\mathrm{i}} }/{\beta}})n^{{\text s}}$ so that the gauge equations (\ref{theta}) reduce to the following form:
\begin{subequations}\label{gauge}
    \begin{align}\label{eq:finalthetaC}
       { \dot\theta^{{\text c}} }(\tau)&=\frac{2\pi}{\beta}N  -\widetilde\varphi^{{\text c}}(\tau),
    \\\label{eq:finalthetaS}
        {\dot\theta^{{\text s}} }(\tau)&=\frac{2\pi{\mathrm{i}} }{\beta}n^{{\text s}} -\widetilde\varphi^{{\text s}}(\tau).
    \end{align}
\end{subequations}

After the gauge transformation the mixed action in Eq.~(\ref{eq:actions}) is reduced to the following quadratic fermionic action in terms of the transformed fields:
\begin{align}\label{eq:psiaction}
 S_{{\text{f}}}=
\int_0^\beta\!\!
\mathrm{d}\tau&\sum_\alpha\widetilde{\overline{\Psi}}_\alpha(\tau) \Big[\partial_\tau+\varepsilon _\alpha -\widetilde{\mu}_\sigma \Big]\!\widetilde{\Psi}_\alpha(\tau)\,
.
\end{align}
The zeroth components of the bosonic fields (\ref{phi0}) enters Eq.~(\ref{eq:psiaction}) via the
spin-dependent effective chemical potential $\mu_\sigma$ given by
\begin{equation}\label{mutilde}
\widetilde{\mu}_\sigma=\mu-\mathrm{i} \varphi^{{\text c}}_0-\sigma\varphi^{{\text s}}_0-\frac{2\pi i}{\beta}\left(N  +\sigma n^{{\text s}} \right),
\end{equation}
where $\sigma=\pm1$ for spin up/down respectively.

The gauge equations (\ref{gauge}) become important for correlation
functions which are not gauge invariant
but depend on phase terms which
are functions of the gauge parameters $\theta^{{\text c}} $ and $\theta^{{\text s}} $ (e.g.\ the Green's function calculated in the following Section and Appendix \ref{sec:GF}).
These parameters are functionals of the
bosonic fields $\widetilde\varphi^{{\text c}}(\tau) $ and $\widetilde\varphi^{{\text s}}(\tau) $ respectively. Thus, in order to
calculate these phase terms, one should solve the gauge
equations and then carry out the integration over the bosonic
fields.  \cite{kamenev1996zba,Grishin}.

Here however, we consider an alternative method, which bypasses the
need to carry out the functional integrals over $\widetilde\varphi^{{\text {c,s}}}(\tau)$. Even though in our case these integrations pose no
great difficulty, the method we consider has general applicability
and could be used in cases where such integrations are impossible to
perform analytically.

Our approach is to view the gauge equations (\ref{gauge})
 as classical Langevin equations governing
the stochastic dynamics of  $\theta^{{\text c}} $ and $\theta^{{\text s}} $,
with the bosonic fields playing the role of noise. The distribution
of the  noise  is determined by the bosonic actions $S^{{\text c}}$ and
$S^{{\text s}}$, Eq.~(\ref{eq:actions}).

The Langevin equations can be mapped, via the standard tools of
classical stochastic analysis\cite{risken}, to Fokker-Planck (FP)
equations from which the time dependent distribution functions for
$\theta^{{\text c}} $ and $\theta^{{\text s}} $ can be determined.
As an example, the form of the FP equation derived from Eq. (\ref{eq:finalthetaC}) is
\begin{equation}\label{eq:FPthetaC}
\frac{\partial\mathcal{P}^{{\text c}} }{\partial\tau}=\left(\frac{2\pi}{\beta}N  -\mathrm{i} \zeta\right)\frac{\partial
\mathcal{P}^{{\text c}} }{\partial\theta}+E_{\text{c}} \frac{\partial^2 \mathcal{P}^{{\text c}} }{\partial\theta^2},
\end{equation}
where $\mathcal{P}^{{\text c}} $ is the distribution function  for the gauge parameter $\theta^{{\text c}} $
and $\zeta$ is a constant (details regarding the
transition from Langevin to FP equations and their solution are
given in Appendix \ref{sec:PCF}). Equation
(\ref{eq:FPthetaC}) is a standard diffusion equation with a drift
term, the solution of which (with an appropriate initial condition)
is simply a decaying Gaussian, explicitly given by Eq.~(\ref{eq:AP}).

This distribution, and a similar one for $\theta_{\text{S}}$, can now be used to calculate the averaging of any
phase terms involving the
gauge parameters in the calculation of
non gauge-invariant correlation functions.    Thus we can, in effect, replace a functional
integration with an integration over a finite number of
parameters. This is  an alternative method by which to integrate
out the finite frequency components of the bosonic fields
$\varphi^{{\text c}}(\tau)$ and $\varphi^{{\text s}}(\tau)$.

\section{Single Particle Green's Function: Effective Charge Quantization}
\label{sec:GCtoC}
We begin with calculating the temperature Green's function (GF) in the grand canonical ensemble and will show that in the Coulomb blockade regime it reduces naturally to one in the canonical ensemble. Our  starting  expression is:
\begin{align}\label{eq:AGF1}
\mathcal{G}_{ \sigma}&(\tau,\,\mu) = \sum_{\alpha} \mathcal{G}_{\alpha,\sigma} (\tau,\,\mu)\,,  \\[4pt]\nonumber
\mathcal{G}_{\alpha,\sigma} &\equiv \frac{1}{\mathcal{Z}(\mu)}
\int\mathcal{D}[\overline{\Psi}_\alpha \Psi_\alpha]\mathrm{e}^
{-S\left[\overline{\Psi}_\alpha\Psi_\alpha\right]}\overline{\Psi}_{\alpha,\sigma} (\tau_i)\Psi_{\alpha,\sigma}(\tau_f),
\end{align}
where $\mathcal{G}_{\alpha,\sigma} $ is an auxiliary GF corresponding to a level $\varepsilon _\alpha$,
$S\left[\overline{\Psi}_\alpha\Psi_\alpha\right]$ is the $\alpha$-term in the Euclidean action  (\ref{eq:action1}) and
$\tau\equiv\tau_f-\tau_i$.

After the HS transformation and gauge transform (\ref{GO}), the Gaussian integration over the quadratic fermionic action (\ref{eq:psiaction}) is straightforward. The resulting GF of non-interacting electrons corresponding to this action, $\mathcal{G}^{0}_{\alpha,\sigma}(\tau,\widetilde{\mu}_\sigma)$, depends -- via Eq.(\ref{mutilde}) -- only on the zero-frequency component of the bosonic fields $\varphi ^{\text{a}}_0$. This allows us to subdivide the remaining functional integration with the bosonic part of the action (\ref{eq:actions}) into that over the zero-frequency, $\varphi ^{\text{a}}_0$, and   finite frequency, $\widetilde\varphi ^{\text{a}}$, components, which results in the following expression:
\begin{align}\label{Av0}
    \mathcal{G}_{\alpha,\sigma} =\Pi^{{\text c}} (\tau)\Pi^{{\text s}} (\tau)\, \frac{\left\langle{\!\left\langle{\mathcal{Z}^{0}(\widetilde{\mu}) \mathcal{G}^{0}_{\alpha,\sigma}(\tau,\widetilde{\mu}_\sigma) }\right\rangle}\!\right\rangle_{\!0} }{\left\langle{\!\left\langle{{\mathcal{Z}^{0}(\widetilde{\mu})} }\right\rangle\!}\right\rangle_0}
\end{align}
Here $\Pi^a({\tau})$ are the phase correlation functions resulting from the functional averaging of the charge or spin phase factors over the finite-frequency components of the appropriate fields, $\left<\!\left<{\dots}\right>\!\right>_0$ stand for the functional integrals over the zeroth-component fields $\varphi ^{\mathrm{c}}_0$ and $\varphi ^{\text{s}}_0$. All these functional integrals are defined in Eq.~(\ref{AGF3})--(\ref{Pi}) in Appendix \ref{sec:GF}. Then    $\mathcal{Z}^{0}(\widetilde{\mu})= \mathcal{Z}^{0}_\uparrow(\widetilde{\mu}_\uparrow)
\mathcal{Z}^{0}_\downarrow(\widetilde{\mu}_\downarrow)$ and $\mathcal{G}^{0}_{\alpha,\sigma}(\tau,\widetilde{\mu}_\sigma)$
is the grand canonical partition function\cite{comGPF} of non-interacting electrons with the spin-dependent chemical potential $\widetilde\mu_{\sigma}$, defined by Eq.~(\ref{mutilde}).

The charging effects can be fully accounted for by introducing winding numbers in the integration over $\varphi^{{\text c}}_0$:
\begin{align}\label{WN}
    \varphi^{{\text c}}_0&=\omega_m+\frac{\widetilde\varphi^{{\text c}}_0}{\beta} \,,& \omega_m&=\frac{2\pi}{\beta}m
\end{align}
where $-\pi<\widetilde\varphi^{{\text c}}_0\leq\pi$ and an integer $m$ is a  winding number.
 In the original work of Gefen and Kamenev
\cite{kamenev1996zba} these were not considered, leading to an
incorrect final result. They were first introduced in the context of
the charging interaction on small metallic grains by Efetov and
Tschersich \cite{efetov2003ceg} within a Matsubara framework, and
were finally correctly implemented by Sedlmayr, Yurkevich and Lerner \cite{SYL:06}
within a Keldysh-technique framework.
  The introduction of the winding numbers (\ref{WN})  allows us to replace  integration over
$\varphi^{{\text c}}_0 $ with summation over all integers
\emph{m}   and integration over
$\widetilde\varphi^{{\text c}}_0$. The sum over $m$ is performed using
the Poisson formula,   which results in a new summation of the form
\begin{align}
\nonumber
\sum_N \mathrm{e}^{-\beta E_{\text{c}} \left(N-N_0\right)^2}\times\mathcal{F}(N).
\end{align}
The Poisson resummation  transforms  summation over $m$ into summation over the conjugate
variable, $N$. In our case $\varphi^{{\text c}}_0$ represents a phase, whose
conjugate is evidently the particle number $N$. While the sum over the parameter $m$ had many
contributions (since $(\beta E_{\text{c}} )^{-1}\ll1$), the sum over $N$
contains, under the conditions (\ref{ineq}), only two terms
$N=N_0\pm\tfrac{1}{2}$, near the Coulomb peak ($N_0$ is half an
integer) and one term in the
Coulomb valleys (i.e.\ everywhere outside of the region of width $T$ near the peak): the contribution of all the other terms is exponentially suppressed.
This is a manifestation of charge quantization
in QDs.

In this way we perform the integration in Eq.~(\ref{Av0}) to find (see Appendix \ref{sec:GF}):
\begin{align}\label{eq:GF6}
&\mathcal{G}_{\alpha,\sigma}(\tau,\mu)=\frac{\widetilde{\Pi}^{{\text c}} (\tau)\widetilde{\Pi}^{{\text s}} (\tau)}
{\widetilde{\mathcal{Z}}(\mu)}\sum_N
\mathrm{e}^{-\beta E_{\text{c}} \left(N-N_0+\frac{\tau}{\beta}\right)^2} \mathcal{I}_N \,, \\&\mathcal{I}_N\equiv\int\limits_{-\infty}^{\infty}\!\!\!
\mathrm d\widetilde\varphi^{{\text s}}_0\, \mathrm{e}^{-\frac{[\widetilde\varphi^{{\text s}}_0]^2}{\beta J}} \!\!\int\limits_{-\pi}^{\pi}\!\!\frac{\mathrm d\widetilde\varphi^{{\text c}}_0}{2\pi}\, \mathrm{e}^{{\mathrm{i}}\left(N+\frac{\tau}{\beta}\right) \widetilde\varphi^{{\text c}}_0}\mathcal{Z}^{0} (\widetilde{\mu})
\mathcal{G}^{0}_{\alpha,\sigma}(\tau,\widetilde{\mu}_\sigma)\,,\label{I0}
\end{align}
where the reduced phase correlation functions $\widetilde\Pi^a$ are defined in Eq.~(\ref{PiTilde}).
The effective charge quantization  in   Eq.~(\ref{eq:GF6}) makes it natural  to change over from grand canonical to canonical
quantities for a given $N$,  followed by a
weighted summation over $N$, where required. Let us stress that the canonical quantities are auxiliary and we calculate in this way the grand canonical GF of Eq.~(\ref{eq:GF6}).

Expressing ${\mathcal{I}}_N$ via canonical quantities leads to an extra summation since ${\mathcal{Z}}=\sum_{n} {\mathrm{e}}^{\beta \mu n}Z_n $, etc. This calculation is detailed in Appendix \ref{ApB}. The resulting full single particle GF in imaginary time (following
summation over all single particle energy states) is given by
 \begin{align}\label{eq:GFfinal11}
\mathcal{G}(\tau,\mu)&
=\frac{\pi T}{\Delta}\frac{\mathrm{e}^{-(E_{\text{c}} -
{J}/{4})|\tau|}
}{\sin\bigl(
{\pi|\tau|}T\bigr)}  \, \frac{F({\tau})}{F({0})}  ,
\end{align}
 where
\begin{align}
 \label{Z}
&F({\tau})=\sum_N\mathrm{e}^{-\beta E_{\text{c}}\left(\delta N\right)^2}\!\!\! \sum_{M=-N}^N\!\!\mathrm{e}^{-\frac{1}{4}\beta\left(\Delta-J \right)M^2 - \tau{E}_{N,M}}\,,\\
\notag
&  \delta N \equiv N- N_0-\frac{\mu}{2E_{\text{c}} },
 \quad
{E}_{N,M} \equiv 2E_{\text{c}}\, \delta N -\frac{JM}{2}\,.
\end{align}
The double summation above arises from replacing the grand canonical partition function in terms of the sum over canonical ones, ${\mathcal{Z}}(\mu)=\sum_{n} {\mathrm{e}}^{ \beta\mu n}Z_n $. The summation parameters are the electron number, $N$, and the total spin of the dot (in the units of ${\hbar}/2$), $M$. Naturally, the GF is  spin independent: we are considering the regime of parameters,  Eq.~(\ref{ineq}), below the Stoner instability where there is no symmetry breaking to distinguish opposite spin polarizations. Note that this result is valid in the regime (\ref{ineq}),    provided that
\begin{align}\label{N}
N&\Delta\gg T\,,&(N-|M|)&\Delta\gg T\,,
\end{align}
i.e.\ when the QD contains many electrons and is not very close to the Stoner instability. Moreover, under these conditions the sum over $M$ in Eqs.~(\ref{eq:GFfinal11}) and (\ref{Z}) can be replaced by an integral from $-\infty$ to $+\infty$ and the exponent of $\tfrac{J^2\tau^2}
{4\beta(\Delta- J)}$   resulting from this integration can be totally neglected. With the same accuracy, we should neglect the exchange energy $J$ in the exponent in Eq.~(\ref{eq:GFfinal11}). Thus we find \begin{align}
\mathcal{G}(\tau,\mu)&=\frac{{\pi}\mathrm{e}^{- E_{\text{c}}  |\tau|}
}{{\beta\Delta}\sin\bigl(\tfrac{\pi|\tau|}{\beta}\bigr)}  \, \frac{1}{\widetilde{Z}}\sum_N \mathrm{e}^{- E_{\text{c}} \left[\beta({ \delta N })^2-2\tau\,\delta N  \right]}\,,
\label{GFtotal}
\end{align}
so that  under  conditions (\ref{ineq}) and (\ref{N})  -- not surprisingly --  the one-particle GF is independent of the exchange part of the universal Hamiltonian (\ref{H}). Such a dependence would emerge only very close to the Stoner instability, when $|\Delta-J|/J\ll1$ but this parametric region is beyond the scope of the presented technique.

\section{Tunneling Density of States}
\label{sec:TDoS}

The tunneling density of states (TDoS), $\nu({\varepsilon })$, can be directly related to the conductance of the QD in the limit of weak coupling to the leads
and is thus a quantity of great importance.  The TDoS is given by $\nu({\varepsilon })=-\frac{1}{\pi}\operatorname{Im}\mathcal{G}^R(\varepsilon )$, where the retarded GF, ${G}^R(\varepsilon )$, is a Fourier transform of the GF in real time, $G({t,\mu})$, obtained from Eq.~(\ref{GFtotal}) by the straightforward analytical continuation from the upper half-plane. Since ${\mathcal{G}}({\tau,\mu})$ is independent of the exchange energy under the conditions (\ref{ineq}) and Eq.~(\ref{N}), so is the TDoS.\cite{SYL:06}

For tutorial purposes, we use the results of Appendices \ref{sec:GF} and \ref{ApB} to derive a more general expression for $\nu({\varepsilon })$, valid for any relation between the parameters in Eqs.~(\ref{ineq}) and (\ref{N}) and show how it goes over to the known expression \cite{SYL:06} under  conditions (\ref{ineq}) and (\ref{N}).

Using the GF in the $\varepsilon $-representation, Eq.~(\ref{GR}), and performing the summation over all the levels as described at the end of Appendix \ref{ApB}
  we find
  \begin{align}\label{TDoS}
\nonumber
&\frac{\nu(\varepsilon)}{\nu_0}=\frac{1}{\widetilde{Z}}\sum_N\sum_{M=-N}^N \mathrm{e}^{-\beta E_{\text{c}} \left(N-\widetilde{N}_0\right)^2 -\frac{1}{4}\beta(\Delta- J)M^2}\times
 \\
&\Big[1-n\left(\varepsilon-\bar\mu-\xi_{N,M}\right)
+n\left(\varepsilon-\bar\mu-\xi_{N\!-\!1,M\!-\!1}\right)\Big],
\end{align}
where we have defined
\begin{equation}\label{xiNM}
\xi_{N,M}\equiv 2E_{\text{c}} (N-\widetilde N_0+\tfrac{1}{2})-\tfrac1{2}{J}(M+\tfrac{1}{2})\,,
\end{equation}
and
\begin{align}\label{mubar}
    \bar\mu&\equiv \tfrac12\Delta(N+M)\,, &\widetilde{N}_0&\equiv N_0+\frac{\mu}{2E_{\text{c}} }\,,
\end{align}
 while $\nu_0={2}/{\Delta}$ is the TDoS in the absence of interactions, $n({\epsilon})\equiv \left[1+\mathrm{e}^{\beta\epsilon}\right]^{-1}$.
\begin{figure}[floatfix]
    \subfigure[]{\includegraphics[width=0.9\columnwidth]{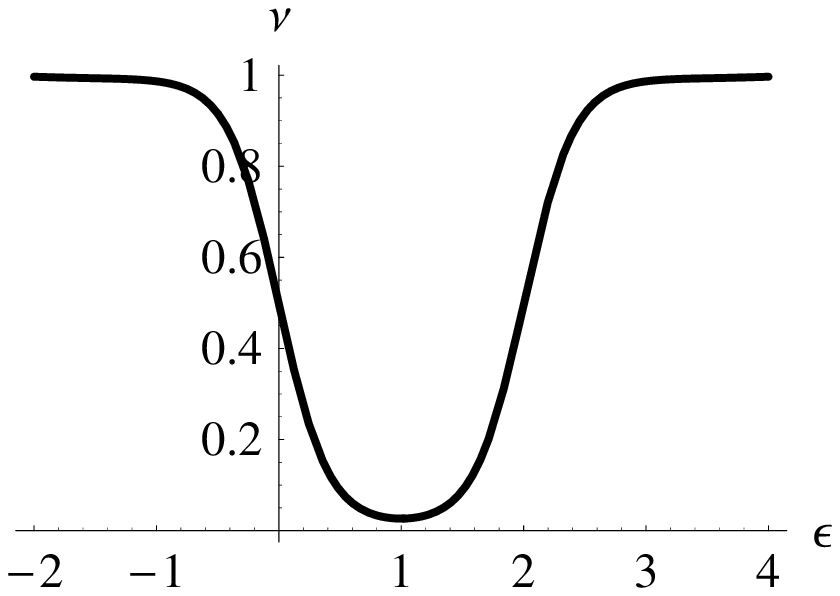}}
    \subfigure[]{\includegraphics[width=0.9\columnwidth]{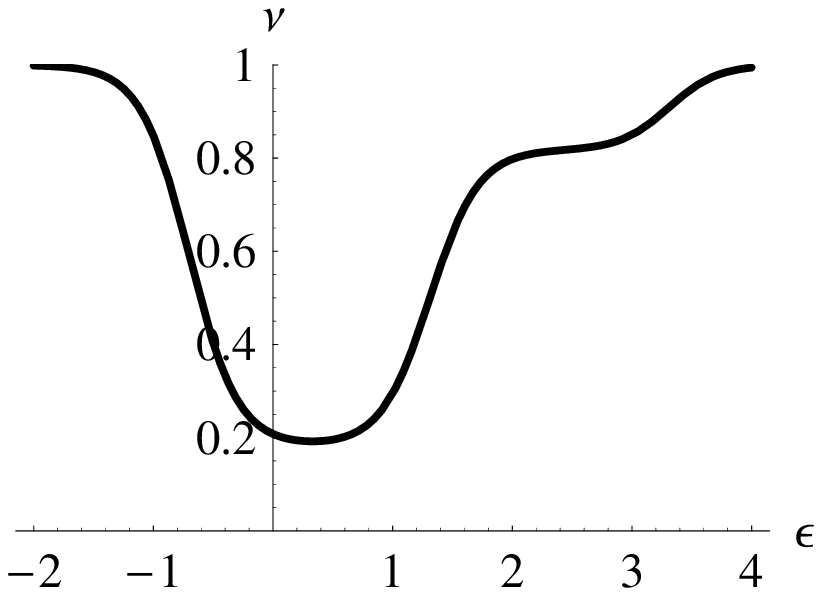}}
    \subfigure[]{\includegraphics[width=0.9\columnwidth]{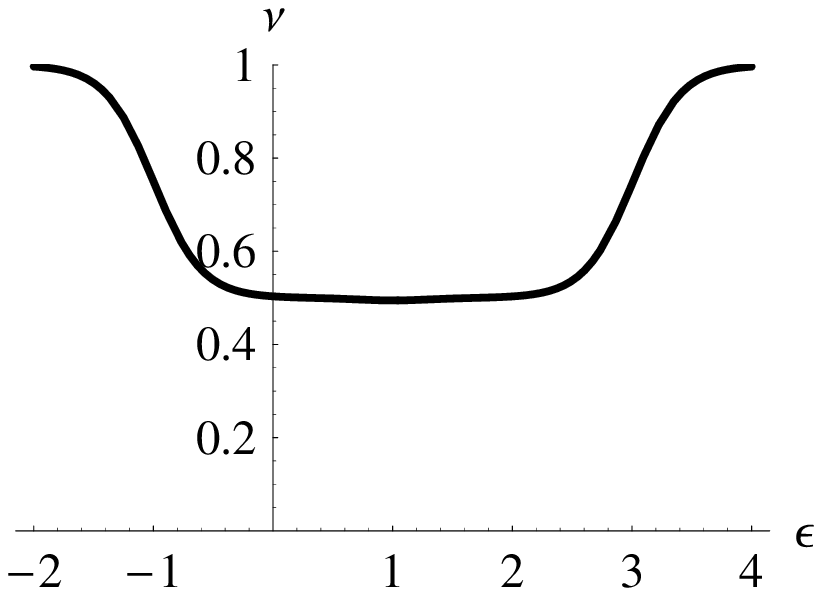}}
  \caption[TDoS as function of energy]{TDoS  (in units of $\nu_{0}$) as a function of
 $\epsilon\equiv\varepsilon /E_{\text{c}} $ for $T=0.2E_{\text{c}} $ and
  ${\Delta}/{T}=0.1$  in (a) a CB valley ($\widetilde{N}_0=100$), (b) an intermediate region ($\widetilde{N}_0=100.35$), (c) a
  CB peak ($\widetilde{N}_0=100.5$).}
  \label{fig:tdos1}
\end{figure}

\begin{figure}[floatfix]
\includegraphics[width=.9\columnwidth]{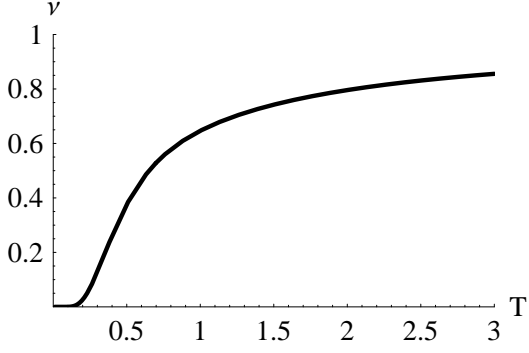}
\caption[TDoS as function of temperature]{Dependence of the TDoS (in units of $\nu_{0}$) on the temperature (measured
in $E_{\text{c}} $) at the bottom of a CB valley ($\widetilde{N}_0=70$) for ${\Delta}/{E_{\text{c}} }=0.02$.}\label{fig:tdos2}
\end{figure}

Equation (\ref{TDoS}) is the general expression for the TDoS for any combination of parameters for a many-electron dot. When the inequalities (\ref{ineq}) and  (\ref{N}) are satisfied, we can easily sum over $M$ as described at the end of the previous section and then limit the summation over $N$ to the two terms for which the value of $|N-\mathcal{N}_0|$ is minimal (although deep in the Coulomb valley only one term is actually contributing). The resulting TDoS is independent of $J$ (or, more precisely, tiny $J$-dependent corrections are beyond the accuracy of current calculations and thus omitted) and coincides with that obtained in Ref.~\onlinecite{SYL:06}:
\begin{align}\label{nuf}
\frac{\nu(\varepsilon)}{\nu_0}=\frac{U(\varepsilon-\xi_N)
+\mathrm{e}^{-\beta(\xi_N-\bar\mu)}\,
U(\varepsilon-\xi_{N+1})}{1+\mathrm{e}^{-\beta(\xi_N-\bar\mu)}} ,
\end{align}
where $
U(\varepsilon-\xi_N)\equiv n(\varepsilon-\xi_{N-1}-\bar \mu)+1-n(\varepsilon-\xi_N-\bar \mu)
$, and $\xi _N$ is obtained from $\xi _{N,M}$ by putting $J=0$ in Eq.~(\ref{xiNM}).
We illustrate the dependence of $\nu$ on energy for integer, half-integer and intermediate values of $\widetilde{N}_0$  in Fig.~\ref{fig:tdos1}, for a specific choice
of parameter values $T $ and $\Delta$. Its dependence on
temperature at the bottom of a Coulomb blockade valley is depicted
in Fig.~\ref{fig:tdos2}. It is important to note that  the TDoS  obtained in the Coulomb valleys is not
physical since we neglect co-tunneling contributions; however, the $T$-dependence near the peak will be obtained as a linear combinations of those shown in Fig.~\ref{fig:tdos2}.

Note that for any given set of parameters the center of the TDoS curve is at $\varepsilon_0=\tfrac{1}{2}\Delta\widetilde{N}_0-2E_{\text{c}} \left(N-\widetilde{N}_0\right)$
and thus a function of $\widetilde{N}_0$, Eq.~(\ref{mubar}). This moving from one Coulomb valley to the next, the TDoS curve is shifted by $\Delta/2$ due to adding an extra electron to the dot,
which raises the effective chemical potential and thus
shift the TDoS curve. That is the reason for the `half-gap' in TDoS at the degeneracy point.

\section{Magnetic Susceptibility}
\label{sec:MS}

We now turn to calculating the longitudinal and transverse magnetic susceptibilities of the system.

It is clear that only the static component of the longitudinal susceptibility is non-zero  due to the lack of spin flip processes in the Ising model.\cite{Note1}
A direct calculation of the correlation function $\langle S_z(\tau)S_z(0)\rangle$ shows this to be $\tau$-independent, as expected. The static susceptibility is given by
\begin{equation}\label{eq:staticLMSdef}
\chi_{zz}=\frac{1}{\beta}\lim_{h\rightarrow0}\frac{\mathrm{d}^2}{\mathrm{d}h^2} \ln\mathcal{Z}(h),
\end{equation}
where $\mathcal{Z}(h)$ is the partition function of the system
calculated in the presence of the following source term in the action:
\begin{align}\label{h}
 S_h= -\frac{h}{2}\int_0^\beta
\mathrm{d}\tau\sum_\alpha\overline{\Psi}_\alpha\sigma^z\Psi_\alpha.
\end{align}
The calculation is straightforward,  leading to the result
\begin{align*}
\mathcal{Z}(h)=\kappa \exp{\left\{\frac{\beta^2 h^2}{4\beta(\Delta-J)}\right\}},
\end{align*}
with $\kappa$ being some irrelevant constant. Plugging this into the definition (\ref{eq:staticLMSdef}) yields
the well known expression
\begin{equation}\label{LMS}
\chi_{zz}({\omega=0}) =
\frac{1}{2}\,\frac{1}{\Delta-J} \,.
\end{equation}
As expected, the static susceptibility is independent of the number of particles on the dot, external gate voltage, charging effects, etc.

We now turn to a calculation of the transverse magnetic susceptibility. This quantity is inherently different from the longitudinal one since it is dynamic: the model allows for transitions between different transverse spin polarization states.

We define the dynamic transverse susceptibility in imaginary time as
\begin{equation}\label{eq:def}
\frac{1}{\beta}\chi^{+-}(\tau)=\langle\sigma^+(0)\sigma^-(\tau)\rangle,
\end{equation}
where
$\sigma^+=\sum_\alpha\overline{\Psi}_{\alpha\uparrow}\Psi_{\alpha\downarrow}$
and
$\sigma^-=\sum_\alpha\overline{\Psi}_{\alpha\downarrow}\Psi_{\alpha\uparrow}$.
Thus we need to calculate the functional average of
\begin{align*} \sum_{\alpha,\beta}\overline{\Psi}_{\alpha\uparrow}(0)
\Psi_{\alpha\downarrow}(0)\overline{\Psi}_{\beta\downarrow}(\tau) \Psi_{\beta\uparrow}(\tau),
\end{align*}
with the action  given by Eq. (\ref{eq:action1}). The  procedure closely follows to that of the calculation of the GF described in Section \ref{sec:GCtoC}. The final outcome
of this calculation is
\begin{align}\label{eq:fintms1}
\chi^{+-}(\tau)=&\frac{\beta\mathrm{e}^{J\tau}}{\widetilde{\mathcal{Z}}(\mu)} \sum_N \mathrm{e}^{-\beta E_{\text{c}} (N-\widetilde{N}_0)^2}\!\!\!\sum_{M=-N}^N \Big\{\mathrm{e}^ {-\frac14{\beta}(\Delta-J)M^2}\times \nonumber\\
&\mathrm{e}^{J\tau M}\sum_\alpha
\left[1-n_\alpha(\bar\mu_\uparrow)\right]n_ \alpha(\bar\mu_\downarrow)\Big\}\,,
\end{align}
where $\bar \mu_\sigma\equiv N_\sigma\Delta$ and $N_\sigma$ is the total number of electrons with the spin projection $\sigma=\uparrow, \downarrow$.

\begin{figure}[fixfloat]
    \subfigure[]{\includegraphics[width=\columnwidth]{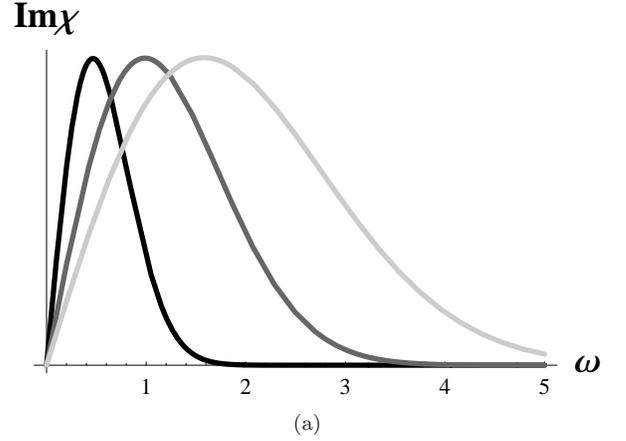}}
    \subfigure[]{\includegraphics[width=\columnwidth]{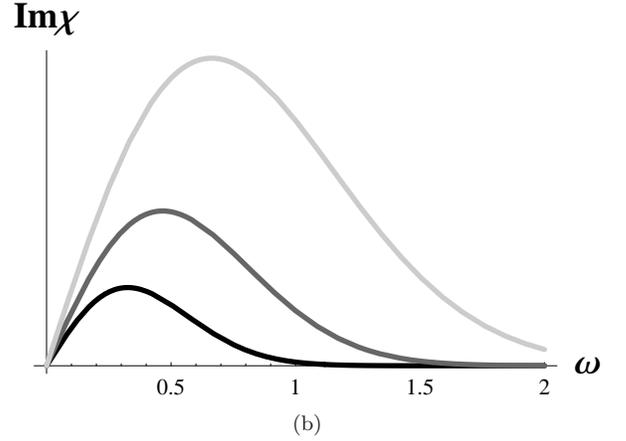}}
  \caption[Transverse Magnetic Susceptibility as function of frequency]{$\frac{1}{\beta}\textrm{Im}\chi^{+-}$
 as a function of frequency $\omega$ (in units of $\Delta$) for (a)
$\tfrac{\Delta}{T}=0.1$ and
  $\tfrac{J}{\Delta}=0.1,0.2,0.3$ for the left, center and rightmost curves respectively and (b) $\tfrac{J}{\Delta}=0.1$ and
  $\tfrac{\Delta}{T}=0.05,0.1,0.2$ for the top, center and bottom curves respectively}\label{fig:sus}
\end{figure}

\begin{figure}[fixfloat]
\begin{center}
\includegraphics[width=\columnwidth]{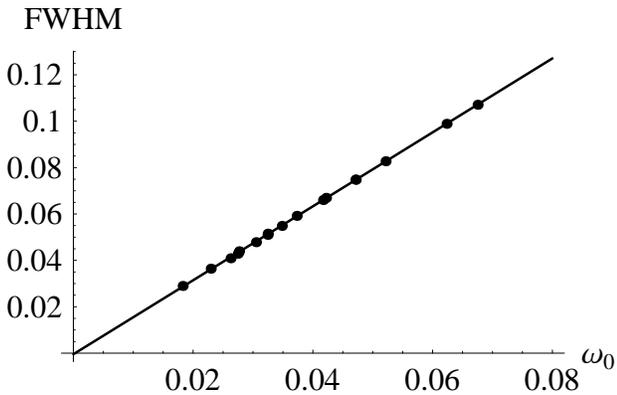}
\caption[Numerical fit of data for full-width-at-half-maximum (FWHM)
of $\frac{1}{\beta}\textrm{Im}\chi^{+-}$]{Fit of numerically
acquired data for FWHM to function $FWHM=\alpha\omega_0$, yielding
$\alpha=1.59$. $R^2$ for this fit is 0.999.}\label{fig:FWHM}
\end{center}
\end{figure}

Fourier-transforming the result of Eq.~(\ref{eq:fintms1})  to Matsubara frequencies and then performing a simple
analytic continuation, we find  the imaginary part of the
physical response function $\chi^{+-}(\omega)$:
 \begin{align}\label{eq:imchi}
\nonumber
 \textrm{Im}\chi^{+-}(\omega)&=\frac{\sqrt{\pi{\beta} (\Delta- J)}}{2
J}\mathrm{e}^{\frac{\beta}{4}\left[(\Delta+J) - (\Delta-J)\frac{\omega^2}{J^2} \right]} \\&\times \left(1+ \frac{\omega}{J}\right)\frac{\sinh\left[\tfrac{\beta\omega}{2}\right]}
{\sinh\left[\tfrac{\beta\Delta}{2}\left(1+\tfrac{\omega}{J}\right)\right]}
\end{align}
This function is depicted in Fig.~\ref{fig:sus}.
The most salient features are a linear dependence at the
origin and the existence of a peak at a certain $\omega_0$. Both the slope at the origin and the value of $\omega_0$ can be used
used to characterize an experimentally obtained curve of the
transverse magnetic susceptibility as a function of frequency.  We find the slope at $\omega\rightarrow0$ as
\begin{align}\label{eq:slope}
\frac{1}{\beta}\textrm{Im}\chi^{+-}(\omega\rightarrow0)
&\approx\frac{\omega}{2J} \sqrt{\frac{\pi}{\beta\Delta}},
\end{align}
where the approximation was made consistent with the inequality (\ref{ineq}). Under the same condition, the peak frequency is given by
\begin{equation}\label{eq:resfreq}
\omega_0\approx\sqrt{\frac{2J^2}{\beta\left(\Delta- {J}\right)}}
\,.
\end{equation}
Yet another parameter of interest is the full-width-at-half-maximum
(FWHM). Numerical analysis shows that it is
proportional to the resonance frequency: $
FWHM\approx1.59\omega_0$.
     This result was
    derived by numerically obtaining the FWHM for various values of
    $\omega_0$ and fitting the results to a linear curve, as shown in Fig.~\ref{fig:FWHM}.

The imaginary part of the susceptibility represents the systems
capacity to absorb and dissipate magnetic energy at a nonzero
frequency. For the static susceptibility only
a  real part is finite. A simple calculation leads to
\begin{align}\label{eq:real2}
\textrm{Re}\chi^{+-}(\omega=0)=\frac{1}{\Delta}\mathrm{e}^ {\frac{\beta}{4}\left(\Delta+J\right)}\approx\frac{1}{\Delta}.
\end{align}
Note that in the limit $J=0$  we recover the well known identity $\chi^{+-}=2\chi_{zz}$ for the static susceptibilities. The real part of $\chi^{+-}$ at finite frequencies can be found either directly or via the Kramers-Kronig relations but we do not present the result here as it has little physical relevance.

As in the case of the longitudinal magnetic susceptibility, it is clear that the
transverse susceptibility is not affected by the charging interaction in the dot.
Once again we see that under conditions (\ref{ineq}) and (\ref{N})
the charge and spin degrees of freedom are effectively decoupled.

\section{Summary}

The main results of this work fall into three basic categories.
These are the single particle GF, the TDoS, and the magnetic
susceptibilities. The results for all three classes of correlation
functions were obtained by means of the functional bosonization
approach combined with the solution of classical stochastic
equations for the bosonic fields. We considered the Ising version of
the {\it Universal Hamiltonian} for description of the interplay
between the spin and charge degrees of freedom in zero-dimensional
systems. Such model is Abelian and therefore does not include the
physics of non-commutative variables. It also does not exhibit the
mesoscopic Stoner instability regime. Nevertheless, the spin-charge
entanglement is present being manifested in e.g. non-trivial AC spin
susceptibility. The model, being a simplified version of the quantum
Universal Hamiltonian model gives qualitatively correct description
of the thermodynamics and transport through nanostructures in the
vicinity of {\it thermodynamic} Stoner Instability point. The
Stochastic Bosonization appears to be very powerful tool for a
treatment of Abelian gauge theories and a promising method for
solving non-Abelian models corresponding to isotropic/anysotropic
quantum limits of the Universal Hamiltonian. The theory of
thermodynamic Stoner Instability and its influence on the transport
through single electron transistor can be tested experimentally in
quantum dot devices and granular systems%
\cite{Exp:Co_in_Pt,*Exp:Co_in_Pt-1,*Exp:Co_in_Pt-2,%
Exp:Fe_in_TransMet,*Exp:Ni_in_Pd,*Mirza,Menon,*Canfield}.

We summarize below the central results and key observations reported
in the paper.

\begin{itemize}
\item {\it Canonical variables and charge quantization.}
In our calculation of the GF, the tools we used and the choices made not only allowed us to carry out a
non-perturbative calculation, but also had physical significance.
The use of functional bosonization and generalized gauge
transformations and the implementation of winding numbers, as well
as the transformation to conjugated variables via the Poisson
re-summation, led us to employ canonical quantities. The latter is a
consequence of strong charging interaction.

\item {\it Regimes of validity.} The transition to canonical quantities, namely the introduction
of the canonical partition function,
also led to further insight with regard to the various physical
regimes the system may be found in. Our calculation of the canonical
partition function itself (and the associated quantity
$\mathcal{Z}_N (\rlap{/}\varepsilon_\alpha)$) imposed limitations on
the physical parameters involved. We found that the system must be
large enough (meaning a large number of electrons), and far below
the Stoner instability point. We had to self consistently assume
that the fluctuations in the systems magnetization were much smaller
than the system size. This corresponds to a requirement that the
system be far from a phase transition point, which in our case is
the SI point.

\item {\it Spin-charge entanglement.} Introduction of the canonical
partition functions led directly to a summation over all possible
values of the magnetization. These are of course limited to $|M|<N$.
Since the number of particles itself is controlled by the charging
interaction when in the CB regime, and the fluctuations of the
magnetization are influenced by the exchange interaction, this can
be seen as a form of coupling between the charge and spin degrees of
freedom. The coupling between the two interaction channels becomes
important as the magnitude of magnetization fluctuations increases,
i.e. as one approaches the SI point. Only then do values of $M$
which approach the system size become accessible and, consequently,
of physical importance. Far below the SI point, the spin-charge
coupling is very weak, and effects of interplay are minimal. Our
calculation of the TDoS showed the exchange interaction to have an
extremely negligible effect. The magnetic susceptibilities in turn
showed no dependence on the charging interaction.

\item {\it Determining $J$ and $\Delta$.} The calculation of the
transverse magnetic susceptibility is,
to our knowledge, a new result, and perhaps the most important in
this work. As we have discussed previously, the importance of this
result is that it provides an experimental method to determine the
values of the parameters $J$ and $\Delta$. Our result is a direct
prediction of the absorption spectrum of the system, and as such
should be amenable to experimental measurement. The various curve
characteristics which we derived, including the slope at
$\omega\rightarrow0$, the location of the resonance frequency and
the FWHM, should in principal, through their dependence on $J$ and
$\Delta$, allow these values to be ascertained from such a
measurement.

\end{itemize}

\begin{acknowledgments}
We acknowledge useful interaction with I.~Burmistrov, I.~Yurkevich
and O.~Zilberberg. In particular, we thank Zeev Schuss for
illuminating discussions concerning stochastic quantization. This
work has been supported by SPP 1285 "Spintronics", Minerva
Foundation, German-Israel GIF, Israel Science Foundation,  EU
project GEOMDISS, and the EPSRC grant  T23725/01. MK acknowledges support of the Einstein Minerva
Center during his visits to WIS.

\end{acknowledgments}

\begin{appendix}

\section{Grand Canonical Single Particle Green's Function}
\label{sec:GF}

In this appendix we present a detailed non-perturbative calculation of the single
particle Green's Function (GF) for our model system
(\ref{H}). The GF itself was used in order to derive the
tunneling density of states (TDoS), but its calculation also serves
to show the methodology used in calculating the various other
quantities considered in this work.

As discussed in section II, a HS transformation is applied,
reducing the
action to the form presented in Eqs.~(\ref{eq:action2}) and
(\ref{eq:actions}). Carrying out the Gaussian integration over the fermionic fields after the gauge transformation (\ref{gauge}), we obtain  the GF as follows:
 \begin{align}\label{AGF3}
 &\mathcal{G}_{\alpha,\sigma}(\tau,\mu)=\Pi^{{\text c}} (\tau)\Pi^{{\text s}} (\tau)\times
\notag\\&\frac{\int\limits_{-\infty}^{\infty}\!\mathrm d\varphi^{{\text c}}_0\;\mathrm{e}^{-S^{{\text c}}_0\left[\varphi^{{\text c}}_0\right]}
\int\limits_{-\infty}^{\infty}\!\mathrm d\varphi^{{\text s}}_0\;\mathrm{e}^{-S^{{\text s}}_0\left[\varphi^{{\text s}}_0\right]}\;
\left[\mathcal{Z}^{0}(\widetilde{\mu}) \mathcal{G}^{0}_{\alpha,\sigma}(\tau,\widetilde{\mu}_\sigma)\right]}
{\int\limits_{-\infty}^{\infty}\!\mathrm d\varphi^{{\text c}}_0\;\mathrm{e}^{-S^{{\text c}}_0\left[\varphi^{{\text c}}_0\right]}
\int\limits_{-\infty}^{\infty}\!\mathrm d\varphi^{{\text s}}_0\;\mathrm{e}^{-S^{{\text s}}_0\left[\varphi^{{\text s}}_0\right]}\; \mathcal{Z}^{0}(\widetilde{\mu})}\,.
\end{align}

Here  $\mathcal{Z}^{0}(\widetilde{\mu})= \mathcal{Z}^{0}_\uparrow(\widetilde{\mu}_\uparrow)
\mathcal{Z}^{0}_\downarrow(\widetilde{\mu}_\downarrow)$ and $\mathcal{G}^{0}_{\alpha,\sigma}(\tau,\widetilde{\mu}_\sigma)$
are the grand canonical partition function\cite{comGPF} and GF of non-interacting electrons with the spin-dependent chemical potential $\widetilde\mu_{\sigma}$, defined by Eq.(\ref{mutilde}).  Both $\mathcal{Z}^{0}$ and $\mathcal{G}^{0}$ are functions of  the zero-Matsubara components $\varphi _0^{\text{c}}$ and $\varphi _0^{\text{s}}$ of the bosonic fields, over which the  integration in Eq.~(\ref{AGF3}) is carried out with
\begin{align}\label{S0}
    S^{{\text c}}_0&=\frac{\beta[\varphi^{{\text c}}_0]^2}{4E_{\text{c}} }-{\mathrm{i}}\beta N_0\varphi^{{\text c}}_0& S^{{\text s}}_0 &=\frac{\beta[\varphi^{{\text s}}_0]^2}{J}\,.
\end{align}
The functional integration over the remaining components of the bosonic fields results in the appearance of the phase correlation functions:
     \begin{align}\nonumber
        \Pi^{{\text c}} (\tau)&=
        \Big\langle \mathrm{e}^{{\mathrm{i}} \left[\theta^{{\text c}} (\tau_f)-\theta^{{\text c}} (\tau_i)\right]} \Big\rangle_{\widetilde\varphi^{{\text c}}}
   \\[-3pt]\label{Pi}
\\[-6pt]\nonumber
     \Pi^{{\text s}} (\tau)&=\Big\langle \mathrm{e}^{\sigma\left[\theta^{{\text s}} (\tau_f)-\theta^{{\text s}} (\tau_i)\right]} \Big\rangle_{\widetilde\varphi^{{\text s}}}\,.
    \end{align}
The functional averaging above is carried out with the weights $\exp[-\widetilde S^{\text{c,s}}]$, where $ \widetilde S^{\text{c,s}}$ are obtained from the appropriate bosonic action in Eq.~(\ref{eq:actions}) by subtracting the zeroth Matsubara components of Eq.~(\ref{S0}).

The calculation of the the correlation functions of Eq.~(\ref{Pi}) is carried out in Appendix \ref{sec:PCF} using the tools of stochastic analysis. The results are:
\begin{align}\nonumber
\Pi^{{\text c}} (\tau)=\mathrm{e}^{-E_{\text{c}} \left(|\tau|-\tfrac{\tau^2}{\beta}\right)}\mathrm{e}^{2 \pi\mathrm{i}N  \tfrac{\tau}{\beta}}&\equiv \widetilde\Pi^{{\text c}} (\tau) \mathrm{e}^{2 \pi\mathrm{i} N   \tfrac{\tau}{\beta}}
 \\[-3pt]\label{PiTilde}
\\[-6pt]\nonumber
\Pi^{{\text s}} (\tau)=\mathrm{e}^{\tfrac{J}{4}\left(|\tau|- \tfrac{\tau^2}{\beta}\right)}\mathrm{e}^{2 \pi\mathrm{i}n^{{\text s}} \tfrac{\tau}{\beta}\sigma^z}& \equiv \widetilde\Pi^{{\text s}} (\tau)\mathrm{e}^{2 \pi\mathrm{i}N   \tfrac{\tau}{\beta}}
 \,.
\end{align}

At this point we introduce the winding numbers, as discussed in section \ref{sec:GCtoC} of the main text.
Following the transition
$\varphi^{{\text c}}_0=\omega_m+\frac{\widetilde\varphi^{{\text c}}_0}{\beta}$, and utilizing
the identities
$\mathcal{Z}^{0}(\mu-i\omega_m)=\mathcal{Z}^{0}(\mu)$ and
$\mathcal{G}^{0}(\tau,\mu-i\omega_m)=\mathrm{e}^{-i\omega_m\tau} \mathcal{G}^{0}(\tau,\mu)$,
we end up with
\begin{align*}
\!\!&\mathcal{G}_{\alpha,\sigma}(\tau,\mu)=\frac{\widetilde{\Pi}^{{\text c}} (\tau)\widetilde{\Pi}^{{\text s}} (\tau)}{\widetilde{\mathcal{Z}}(\mu)} \sum_{m}\mathrm{e}^{ 2\pi {\mathrm{i}}\left(N_0-\frac{\tau}{\beta}\right)m -\frac{\pi^2 m^2}{\beta E_{\text{c}} }}\,\mathcal{I}_m\,,
\\
\!\!&\mathcal{I}_m\equiv\!\! \int\limits_{-\infty}^{\infty}\!\!
\mathrm d\widetilde\varphi^{{\text s}}_0\!\!\int\limits_{-\pi}^{\pi}\!\! \mathrm d\widetilde\varphi^{{\text c}}_0\,\mathrm{e}^{{-\frac{[\widetilde\varphi^{{\text s}}_0]^2}{\beta J}}-\frac{[\widetilde\varphi^{{\text c}}_0]^2}{4\beta E_{\text{c}} }+
\widetilde\varphi^{{\text c}}_0(\mathrm{i}{N_0-\frac{\pi m}{\beta E_{\text{c}}} })} \mathcal{Z}^{0} (\widetilde{\mu})
\mathcal{G}^{0}_{\alpha,\sigma}\,.
\end{align*}
The grand partition function $\widetilde{\mathcal{Z}}(\mu)$ above is represented by the same double-integral and sum with $\mathcal{G}^{0}$ replaced by $1$. The exponential factors involving $N  $ and $n^{{\text s}} $ arising from the
phase correlation functions and the non-interacting GF cancel each
other out exactly. This is hardly surprising  as they are completely arbitrary.

The summation over the winding numbers above can be performed using the Poisson formula
\begin{equation}
\sum_{k=-\infty}^\infty f(2\pi
k)=\frac{1}{2\pi}\sum_{m=-\infty}^\infty\int_{-\infty}^\infty
\mathrm{e}^{{\mathrm{i}}mx}f(x)\mathrm{d}x\,.
\end{equation}
This results in the expression for GF given by Eqs.~(\ref{eq:GF6}) and (\ref{I0})
in the main text.

\section{Calculations in Auxiliary Canonical Ensemble}
\label{ApB}

We express $\mathcal{Z}^{0}(\mu)$ in  Eqs.~(\ref{eq:GF6}) and (\ref{I0}) via the sum of the canonical partition functions for a system
of $n$   non-interacting   electrons, ${Z}^{0}_n$, using the standard relation
\begin{align}\label{eq:Agctoc1}
\mathcal{Z}^{0}(\mu)=\prod_\alpha\left[1+\mathrm{e}^ {-\beta\left(\varepsilon_\alpha-\mu\right)}\right]= \sum_n \mathrm{e}^{\beta\mu
n}{Z}^{0}_n \,,
\end{align}
To express the results of further integration in a convenient way, we also define the grand canonical and canonical partition functions with one level, $\varepsilon _\alpha$, excluded:
\begin{align}\label{Z-exc}
\mathcal{Z}^{0}(\rlap{/}\varepsilon_\alpha, \mu)=\prod_{\alpha'\ne\alpha}\left[1+\mathrm{e}^ {-\beta\left(\varepsilon_\alpha'-\mu\right)}\right]= \sum_n \mathrm{e}^{\beta\mu
n}{Z}^{0}_n ({\rlap{/}\varepsilon_\alpha})\,.
\end{align}

 Then we  substitute into  Eq.~(\ref{I0}) the finite temperature GF of non-interacting fermions
\begin{equation}\label{eq:AnonintGF}
\mathcal{G}_{\alpha,\sigma}^{0}(\tau>0,\mu)= \mathrm{e}^{-(\varepsilon_\alpha-\mu)\tau} \left(1-n_{\alpha,\sigma}(\mu)\right) ,
\end{equation}
where $n_{\alpha,\sigma}(\mu)$ is the  Fermi-Dirac
 occupation factor.   We limit the
 calculation to $\mathcal{G}(\tau>0)$, since $\mathcal{G}(\tau)=-\mathcal{G}(\tau+\beta)$.
Recalling that $\mathcal{Z}^{0}(\widetilde{\mu})= \mathcal{Z}^{0}_\uparrow(\widetilde{\mu}_\uparrow)
\mathcal{Z}^{0}_\downarrow(\widetilde{\mu}_\downarrow)$   we  cast  Eq.~(\ref{I0}) into the form
\begin{align*}
&\mathcal{I}_N=\int_{-\infty}^\infty \mathrm d\widetilde\varphi^{{\text s}}_0\, \mathrm{e}^{-\frac{[\widetilde\varphi^{{\text s}}_0]^2}{\beta J}}\int_{-\pi}^\pi\frac{\mathrm d\widetilde\varphi^{{\text c}}_0}{2\pi}\,
\mathrm{e}^{{\mathrm{i}}N\widetilde\varphi^{{\text c}}_0}
\mathrm{e}^{-\left(\varepsilon_\alpha-\mu+\sigma\frac{\widetilde\varphi^{{\text s}}_0}{\beta}\right)\tau}\\\times
&\sum_{m,n}\mathrm{e}^{\left[\beta\mu(m+n)-{\mathrm{i}} \widetilde\varphi^{{\text c}}_0(m+n)-\sigma(m-n)\widetilde\varphi^{{\text s}}_0\right]}
\mathcal{Z} _{\sigma,m}(\rlap{/}\varepsilon_\alpha)\mathcal{Z} _{-\sigma,n}
.
\end{align*}
Carrying out the integration over $\widetilde\varphi^{{\text c}}_0$  yields a Kr\"{o}necker delta $\delta_{N,n+m}$. Performing the Gaussian integration over $\widetilde\varphi^{{\text s}}_0$ and defining $M=m-n$ we find
\begin{equation}\notag
\mathcal{I}_N=\!\!\sum_{M=-N}^N \mathrm{e}^{\beta\mu N +\frac{1}{4}\beta J\left(M+^\tau\!/\!_\beta\right)^2-\left(\varepsilon_\alpha- \mu\right)\tau}
{Z}_{\frac{N+M}{2}} (\rlap{/}\varepsilon_\alpha) {Z} _{\frac{N-M}{2}}
 .
\end{equation}
Substituting this into Eq. (\ref{eq:GF6}) yields after straightforward algebraic manipulations
\begin{align}\notag
&\mathcal{G}_{\alpha,\sigma}(\tau>0,\mu)=\frac{1}{\widetilde {\mathcal{Z}}(\mu)} \sum_N\sum_{M=-N}^N
{Z}_{\frac{N+M}{2}} (\rlap{/}\varepsilon_\alpha) {Z} _{\frac{N-M}{2}}\\&\times\mathrm{e}^{-\beta
E_{\text{c}} \left(N-N_0\right)^2+\beta\mu N+\frac{1}{4}\beta JM^2
-(\varepsilon _\alpha+ \xi_{N,M}) \tau},\label{eq:AGF7}
\end{align}
where  $\xi _{N,M}$ are defined in Eq.~(\ref{xiNM}).

The canonical partition functions
$ {Z}  _N$ and $  {Z} _N(\rlap{/}\varepsilon_\alpha)$ are evaluated in Appendix \ref{sec:CPF}, resulting in
\begin{align}
\label{can}
{Z}_N &= \mathrm{e}^{-\frac{1}{2}\beta\Delta
N^2}\,, &
{Z}_N (\rlap{/}\varepsilon_\alpha)&=
\left[1-n_\alpha(\bar{\mu}_0)\right]{Z}_N \,,
\end{align}
where the Fermi factor for the $\alpha^{{\text{th}}}$ level,  $n_\alpha({\bar\mu_0})\equiv \big[1+{\mathrm{e}}^{\beta({\varepsilon  _\alpha-\bar \mu_0})} \big]^{-1}$, is taken with the auxiliary chemical potential $\bar \mu_0\equiv N\Delta$.

Substituting Eq.~(\ref{can}) into Eq.~(\ref{eq:AGF7}), we find:
\begin{align}\notag
&\mathcal{G}_\alpha(\tau>0,\mu)=\frac{1}{\widetilde Z}\sum_N \mathrm{e}^{-\beta E_{\text{c}} \left(N-\widetilde{N}_0\right)^2}\times\\&\sum_{M=-N}^N\!\!
\mathrm{e}^{-\frac1{4}\beta\left(\Delta-J \right)M^2}
\!\Big[1-n_\alpha\left(\bar{\mu} \right)\Big]\mathrm{e}^{- \xi_\alpha(N,M) \tau},
\label{eq:AGFfinal}\end{align}
where $\bar\mu$ and $\widetilde{N}_0$ are defined in Eq.~(\ref{mubar})
and the auxiliary partition function $\widetilde{{Z}}\equiv F({0}) $  is given by Eq.~(\ref{Z}) in the main text. The result is naturally spin-independent. Technically, the formal spin dependence vanished when calculating the integral $\mathcal{I}_N$, Eq.~(\ref{I0}). The GF for negative $\tau$ can be obtained from Eq.~(\ref{eq:AGFfinal}) using $\mathcal{G}(-\tau)=-\mathcal{G}(\beta-\tau)$.

Now we find the full GF  by summing over all single
particle states $\varepsilon _a$. This summation is carried out in the usual way by
making the substitution
$\sum_\alpha\mathcal{G}_\alpha\rightarrow {\Delta}^{-1}\int\limits_0^\infty\mathcal{G}(\varepsilon_\alpha) {\mathrm{d}}\varepsilon_\alpha  $, i.e.\
effectively by averaging over disorder by introducing the mean level spacing $\Delta$. This leads to Eq.~(\ref{eq:GFfinal11}) in the main text.

Finally,  we write the GF in the energy representation. Making the standard  analytical
continuation to the real time, $\tau\rightarrow {\mathrm{i}}t$, and Fourier transforming the GF to
the energy domain we obtain the retarded GF used in the calculation of the TDoS as follows:
\begin{align}\notag
&{G}_\alpha^R(\varepsilon)=\frac{1}{\widetilde{Z}} \sum_N\sum_{M=-N}^N
\mathrm{e}^{-\beta E_{\text{c}} \left(N-\widetilde{N}_0\right)^2 -\frac{1}{4}\beta(\Delta- J)M^2}
\\[-9pt]\label{GR}\\\notag&\times
\left[\frac{1-n_\alpha(\bar{\mu})}{\varepsilon- \varepsilon _\alpha-\xi_{N,M}  + {\mathrm{i}}0}+
\frac{n_\alpha(\bar{\mu})}{\varepsilon-\varepsilon _\alpha -\xi_{N\!-\!1,M\!-\!1} +{\mathrm{i}}0}\right].
\end{align}

\section{ Phase Correlation Functions and Stochastic Analysis}\label{sec:PCF}

Here we use stochastic analysis to calculate the phase correlation function  $\Pi^{{\text{c}} }$
defined in Eq. (\ref{Pi}).   $\Pi^{{\text s}} $ has been calculated in exactly
the same manner.

We note that the gauge equation (\ref{eq:finalthetaC}) can be viewed as a Langevin equation wherein the field $\widetilde\varphi^{{\text c}}(\tau)$ plays the role of the
stochastic force (noise), the distribution of which is governed by the action $S_{\widetilde\varphi^{{\text c}}}$ obtained from the appropriate bosonic action in
Eq. (\ref{eq:actions}) by subtracting the zeroth Matsubara components of Eq.~(\ref{S0}). The noise correlation function is given by
\begin{align}\label{noise}
\langle\widetilde\varphi^{{\text c}}(\tau) \widetilde\varphi^{{\text c}}(\tau')\rangle & =   2E_{\text{c}} \left[\delta\left(\tau-\tau'\right)-\frac{1}{\beta}\right]\,,
\end{align}
which  follows from the expansion of $\widetilde\varphi^{{\text c}}(\tau)$ in terms of Matsubara components:
$\widetilde\varphi^{{\text c}}(\tau)=\sum_{m\neq0}\widetilde\varphi^{{\text c}}_m\mathrm{e}^{-{\mathrm{i}} \omega_m\tau}$. Indeed, the
functional distribution of $\widetilde\varphi^{{\text c}}_m$ is
\begin{align} \notag
&\int\mathcal{D}[\widetilde\varphi^{{\text c}}(\tau)]\mathrm{e}^{-\int_0^\beta
\mathrm{d}\tau\left\{\widetilde\varphi^{{\text c}}(\tau)\left[4E_{\text{c}} \right]^{-1}\widetilde\varphi^{{\text c}}(\tau)\right\}}=\nonumber\\=
&\int\!\! \prod_{m\neq0}\!\!\mathrm d\widetilde\varphi^{{\text c}}_m \,
\mathrm{e}^{-\sum_{m,n\neq0}\widetilde\varphi^{{\text c}}_m\left[\frac{\beta\delta_{m,-n}}{4E_{\text{c}} }\right]\widetilde\varphi^{{\text c}}_n}\,,
\end{align}
which corresponds to
$
\langle\widetilde\varphi^{{\text c}}_m\widetilde\varphi^{{\text c}}_n\rangle= {2E_{\text{c}} }{\beta}^{-1}\delta_{m,-n}
$,  immediately leading to Eq.~(\ref{noise}).

It is convenient to represent the noise field as $\widetilde\varphi^{{\text c}}(\tau)=\eta(\tau)+{\mathrm{i}} \zeta$ with $\eta(\tau)$ a random function and $\zeta$
a Gaussian random variable satisfying
$\langle\widetilde{\eta}(\tau)\rangle= \langle\zeta\rangle=\langle\widetilde{\eta}(\tau)\zeta\rangle=0$,
$\langle\widetilde{\eta}(\tau)\widetilde{\eta}(\tau')\rangle=2E_{\text{c}} \delta\left(\tau-\tau'\right)$ and
$\langle\zeta^2\rangle= {2E_{\text{c}} }/{\beta}$. As $\eta(\tau)$
is standard white noise, we follow the standard procedure \cite{risken} to map the Langevin equation (\ref{eq:finalthetaC})   to a Fokker-Planck (FP) equation:
\begin{equation}\label{eq:FP}
\frac{\partial\mathcal{P}^{{\text c}}_\zeta }{\partial\tau}=\left(\frac{2\pi}{\beta}N  -\mathrm{i} \zeta\right)\frac{\partial
\mathcal{P}^{{\text c}}_\zeta }{\partial\theta}+E_{\text{c}} \frac{\partial^2 \mathcal{P}^{{\text c}}_\zeta }{\partial\theta^2}\,.
\end{equation}
Here $\mathcal{P}^{{\text c}}_\zeta (\theta,\tau;\theta'\!,\tau')$ is the \emph{conditional} transition probability function for a given $\zeta$, formally defined by
$\mathcal{P}^{{\text c}}_\zeta (\theta,\tau;\theta'\!,\tau')=\Big\langle\delta \left[\theta(\tau)-\theta\right]\delta\left[\theta(\tau')-\theta'\right]
\Big\rangle_\eta$ where the $\langle{\dots}\rangle_\eta $ means averaging over the white noise $\eta(\tau)$. The full transition
probability function $\mathcal{P}^{{\text c}} (\theta,\tau;\theta'\!,\tau')$ is given by the subsequent averaging over the quenched (i.e.\ $\tau$-independent) variable $\zeta$ (as, e.g., in Ref.~\onlinecite{KLY}):
\begin{align}\label{PDF}
    \mathcal{P}^{{\text c}} (\theta,\tau;\theta'\!,\tau')=\Big\langle\delta \left[\theta(\tau)-\theta\right]\delta\left[\theta(\tau')-\theta'\right]
\Big\rangle_{\eta,\zeta}\,,
\end{align}
i.e.\ $   \mathcal{P}^{{\text c}} \equiv {\langle\mathcal{P}^{{\text c}}_\zeta \rangle}_\zeta$.

Equation (\ref{eq:FP}) is a standard diffusion equation with a drift term. Its solution, with the natural boundary condition  $\mathcal{P}^{{\text c}} (\theta,\tau;\theta'\!,\tau|\zeta)=\delta(\theta-\theta')$, is  a decaying Gaussian:
\begin{equation}\label{eq:AP}
\mathcal{P}^{{\text c}} (\theta,\tau;\theta',\tau'|\zeta)=\frac{\mathrm{exp}\left\{-\frac {\left[\left(\theta-\theta'\right)+\left(\frac{2\pi}{\beta}N  -\mathrm{i} \zeta\right)|\tau-\tau'|\right]^2}{4E_{\text{c}} |\tau-\tau'|^{\phantom{\int}}}\right\}}{\sqrt{4\pi
E_{\text{c}} |\tau-\tau'|^{\phantom{\frac12\!\!\!}}}}\,.
\end{equation}

Now  we   write $\Pi^{{\text c}} (\tau)$, defined in Eq.~(\ref{Pi}), in terms of the transition probability function (\ref{PDF}):
\begin{align*}
\Big\langle \mathrm{e}^{{\mathrm{i}} \left[\theta^{{\text c}} (\tau_f)-\theta^{{\text c}} (\tau_i)\right]}
\Big\rangle_{\widetilde\varphi^{{\text c}}}=\int\limits_{-\infty}^{\infty}\!\!{\mathrm{d}} \theta
{\mathrm{d}} \theta'\,\mathcal{P}^{{\text c}} (\theta,\tau_i;\,\theta'\!,\tau_f)\mathrm{e}^{-{\mathrm{i}} (\theta-\theta')}\,.
\end{align*}
Substituting here the solution (\ref{eq:AP}), we find the conditional (for a given $\zeta$) phase correlation function as
\begin{align}\label{eq:pcfCfinal}
\Pi^{{\text c}} _\zeta(\tau)= \mathrm{e}^{-E_{\text{c}} \tau}
\mathrm{e}^{-\zeta\tau} \mathrm{e}^{-{\mathrm{i}}  2\pi N  \tfrac{\tau}{\beta}}\,,
\end{align}
where we defined $\tau=|\tau_f-\tau_i|$. Finally, the averaging over the quenched random variable $\zeta$ results in the first of Eqs.~(\ref{PiTilde}). The second one, for $\Pi^{{\text{s}} }({\tau})$, has obtained by applying, step by step, exactly the same procedure.

\section{Calculation of the Canonical Partition Function}
\label{sec:CPF}

In this appendix we evaluate the canonical partition functions
${Z}_N $ and ${Z}_N (\rlap{/}\varepsilon_\alpha)$
defined in Eqs. (\ref{eq:Agctoc1}) and (\ref{Z-exc}). It follows from Eq.~(\ref{Z-exc}) that
\begin{align}\label{eq:canondef}
{Z}_N (\rlap{/}\varepsilon_\alpha)&=\int_{-\pi}^\pi\frac{\mathrm d\varphi^{{\text c}}}{2\pi}\,\mathrm{e}^{{\mathrm{i}}N\varphi^{{\text c}}}{Z}^{0}_\alpha
(\mu= {-\mathrm{i} \varphi^{{\text c}}}/\beta)\nonumber\\&=\int_{-\pi}^\pi\frac{\mathrm d\varphi^{{\text c}}}{2\pi}\,\mathrm{e}^{{\mathrm{i}}N\varphi^{{\text c}}}\,\prod\limits_{\alpha'\neq\alpha}
\left(1+\mathrm{e}^{-\beta\varepsilon_{\alpha'}-\mathrm{i} \varphi^{{\text c}}}\right).
\end{align}
We calculate ${Z}_N (\rlap{/}\varepsilon_\alpha) $  (and thus ${{Z}}_N$) in the saddle-point approximation:
 \begin{align}\label{SP}
 {Z}_N (\rlap{/}\varepsilon_\alpha) \approx&
\mathrm{e}^{-\mathcal{S}_\alpha(\varphi^{{\text c}}_0)}\int_{-\pi}^\pi\frac{\mathrm d\varphi^{{\text c}}}{2\pi}\,\mathrm{e}^{-\frac{1}{2}
\left[ \mathcal{S}''_\alpha( \varphi^{{\text c}}_0)\right]\left(\varphi^{{\text c}}-\varphi^{{\text c}}_0\right)^2}\,,
\end{align}
where
\begin{align}\label{Sa}
\mathcal{S}_\alpha(\varphi^{{\text c}})&=-{\mathrm{i}}N\varphi^{{\text c}}-\ln \prod\limits_{\alpha'\neq\alpha}
\left(1+\mathrm{e}^{-\beta\varepsilon_{\alpha'}-\mathrm{i} \varphi^{{\text c}}}\right).
\end{align}
The saddle-point equation, $ {\mathcal{S}}'_\alpha=0 $, is convenient to write by replacing
$\sum_{\alpha'} f(\varepsilon_{\alpha'})$ with $ {\Delta}^{-1}\int_0^\infty
{\mathrm{d}} \varepsilon f(\varepsilon)$  as in Appendix (\ref{ApB}).  This gives, after  calculating the integral, the following equation for finding $\varphi^{{\text c}}_0$:
 \begin{align}\label{sp2}
  N+ \frac{1}{1+\mathrm{e}^{ \beta\varepsilon_\alpha+\mathrm{i} \varphi^{{\text c}}_0}} =\frac{1}{\beta\Delta}\ln\left(1+\mathrm{e}^{-\mathrm{i} \varphi^{{\text c}}_0}\right)\,.
\end{align}
The Fermi-factor there, being of order $1$, can be neglected, which means that the same saddle-point we would find in a calculation of ${Z}_N $: for large enough $N$  the saddle-point is unaltered by the exclusion of a single   state. Assuming also that $N$ is so large that $\beta N \Delta\gg1$, we find from Eq.~(\ref{sp2}):
\begin{align}\label{phi0}
 -\mathrm{i} \varphi^{{\text c}}_0
 =\beta N\Delta\equiv\beta\bar{\mu}.
\end{align}
In the same approximation $S_\alpha({\varphi^{\text{c}} }_0)=1/({\beta\Delta})$, so that calculating the Gaussian integral in Eq.~(\ref{SP}) gives
\begin{align}\label{ZN}
    Z_N={\mathrm{e}}^{-\frac{1}{2}\beta \overline{\mu}N}\,,
\end{align}
while ${Z}_N (\rlap{/}\varepsilon_\alpha)$ differs only by the exclusion of the level $\alpha$:
\begin{align}\label{ZNa}
    {Z}_N (\rlap{/}\varepsilon_\alpha)=\left[1-n_\alpha(\bar{\mu})\right]{Z}_N\,.
\end{align}

\end{appendix}


\begin{thebibliography}{23}
\bibitem [{com()}]{comGPF}%
  \BibitemOpen
  \href@noop {} {}\bibinfo {note} {The grand partition function can be
  represented as the product of two separate partition functions for spin up
  and spin down particles as the model does not allow for any spin flip
  processes and thus the two spin polarizations are not dynamically
  coupled.}\BibitemShut {Stop}%
\end{thebibliography}
\end{document}